\documentclass[conference]{IEEEtran}
\usepackage{cite}      % Written by Donald Arseneau
\usepackage[usenames,dvipsnames,svgnames,table]{xcolor}
\usepackage{subfigure}
\usepackage{graphicx}  % Written by David Carlisle and Sebastian Rahtz
\usepackage{psfrag}    % Written by Craig Barratt, Michael C. Grant,
\usepackage{url}       % Written by Donald Arseneau
\usepackage{tikz}

\usepackage{amsmath}   % From the American Mathematical Society
\usepackage{amssymb}
\usepackage{algorithm}
\usepackage{algpseudocode}
\usepackage{pifont}
\newcommand{\shorten}[1]{}

\newcommand{\signed}%
    {{\unskip\nobreak\hfill\penalty50
      \hskip2em\hbox{}\nobreak\hfil $\blacksquare$
      \parfillskip=0pt \finalhyphendemerits=0 \par}}

\begin{document}

\title{5G Network Slice Isolation with WireGuard and Open Source MANO: A VPNaaS Proof-of-Concept}

% avoiding spaces at the end of the author lines is not a problem with
% conference papers because we don't use \thanks or \IEEEmembership
\author{\IEEEauthorblockN{Simen Haga, Ali Esmaeily, Katina Kralevska, and Danilo Gligoroski}
\IEEEauthorblockA{
Dep. of Information Security and Communication Technology, Norwegian University of Science and Technology (NTNU)\\
Email: \{simehag, ali.esmaeily, katinak, danilog\}@ntnu.no
}
}
\maketitle

\begin{abstract}
The fifth-generation (5G) mobile networks aim to host different types of services on the same physical infrastructure. Network slicing is considered as the key enabler for achieving this goal. Although there is some progress in applying and implementing network slicing in the context of 5G, the security and performance of network slicing still have many open research questions. In this paper, we propose the first OSM-WireGuard framework and its lifecycle. We implement the WireGuard secure network tunneling protocol in a 5G network to provide a VPN-as-a-Service (VPNaaS) functionality for virtualized network functions. We demonstrate that OSM instantiates WireGuard-enabled services up and running in 4 min 26 sec, with potential the initialization time to go down to 2 min 44 sec if the operator prepares images with a pre-installed and up-to-date version of WireGuard before the on-boarding process. We also show that the OSM-WireGuard framework provides considerable enhancement of up to 5.3 times higher network throughput and up to 41\% lower latency compared to OpenVPN. The reported results show that the proposed framework is a promising solution for providing traffic isolation with strict latency and throughput requirements.

\end{abstract}
%\vspace{-0.2cm}


% for peer review papers, inserts a page break and creates the second title.
% Will be ignored for other modes.
\IEEEpeerreviewmaketitle

\section{Introduction} \label{intro}

As the 5G architecture is still evolving, the security threat landscape of 5G has grown enormously due to the unprecedented increase in types of services and the number of devices \cite{7846998}. 5G is expected to deliver simultaneous services with different latency, throughput, connectivity, and security requirements. These services should be provided on the same infrastructure with the help of network slicing. The key feature of network slicing is the capability to virtualize the underlying infrastructure and create independent logical networks \cite{gligoroski2019expanded}. It follows Software-Defined Networking (SDN) \cite{jammal2014softwaredefined,OMNeTSummit2019} and Network Function Virtualization (NFV) \cite{7243304} principles by decoupling the network functions, such as firewalls, load-balancers, proxy servers, intrusion detectors, and others, from proprietary hardware appliances and running them as software in virtual machines (VMs). These different functions are called Virtualized Network Functions (VNFs). %SDN and NFV provide programmability of the network and enable flexible placement and configuration of virtualized functions. 

\textit{Service function chaining} is a technique for selecting and steering data traffic flows through various network functions, i.e., it interconnects the VNFs in a specific order via virtual links to provide a complete end-to-end service. 
The VNFs, network services, and network slices run on top of the NFV Infrastructure (NFVI). ETSI has done a significant amount of work for monitoring and orchestration of VNFs \cite{OSM:SF:WP}. The high-level NFV management framework, proposed by ETSI, has three functional blocks:
\begin{itemize}
    \item The \textit{Infrastructure} block, composed of the NFVI and Virtualized Infrastructure Manager (VIM), is responsible for providing the virtualization environment for the VNFs. The VIM performs the lifecycle management of virtual resources, such as virtual machines, storage, networking, and connectivity.
    \item The \textit{VNFs, network services, and network slices} block includes the collection of VNFs, and the composition of VNFs into network services to form network slices.  
    \item The \textit{Management and Orchestration (MANO)} component, which controls the lifecycle of the VNFs, network services, and network slices, including their configuration and monitoring.
\end{itemize}

In 5G networks, there will be a portfolio of isolation technologies available rather than a single technology like a Virtual Private Network (VPN). Thus, it will be necessary to integrate and manage a variety of isolation mechanisms on different levels. Moreover, there are different levels of slice isolation: isolation of traffic, isolation of bandwidth, isolation of processing, and isolation of storage \cite{8104638,KOTULSKI2020107135}. References \cite{8104638,KOTULSKI2020107135} outline only the challenges of providing slice isolation without proposing solutions. In particular, reference \cite{8104638} lists several potential slice isolation technologies such as tag-based slices isolation, VLAN-based network slices isolation, VPN-based slices isolation with IPsec, Secure Socket Layer/Transport Layer Security (SSL/TLS), Secure Socket Tunneling Protocol (SSTP), SSH (Secure Shell) and SDN-based isolation. 

We focus on a VPN solution for \textit{isolation of traffic}, meaning that the data flow of one slice cannot move to and be accessed from another slice, although they are using the same VNF(s).
We even further extend the concept of traffic isolation to security isolation, i.e., the property that an unauthorized party outside the slice, for instance, another user of the same infrastructure but belonging to another slice, cannot modify or even eavesdrop the traffic flow of the slice \cite{8377166}. This also provides confidentiality and integrity protection of the tenant's traffic even against the Mobile Network Operator (MNO). 
For instance, a tenant that needs a network fit for a specific use case rents a slice from the infrastructure provider. The network slice is then used by the tenant to provide a network service to the end-users. There is no other way to guarantee confidentiality unless the data is encrypted.

\shorten{
Encrypting the data between chained VNFs can lead to telecommunication operators being more open towards multi-vendor NFV solutions \cite{whitestack:sc:1}. There are two main reasons for telecommunication operators being hesitant to multi-vendor NFV solutions. First, the NFV implementations are not steered by operators, but rather by the vendors and integrators that perform the original root cause analysis of the problems. Here the risk lies in the fact that most vendors' business rely on proprietary boxes, which could be negatively impacted by moving to a pure NFV platform. Second, operators are used to managing contracts with a handful of large vendors that provide the design, engineering and network deployment. Thus, the implementation of a multi-vendor NFV solution brings unwanted complexity for operators with regards to whom to blame for security issues. For this reason, most operators are adopting a vertical NFV solution, where a vendor delivers the full NFV deployment and retains the responsibility. 
A horizontal NFV model is the most preferred solution in the telecommunication community, as it will significantly improve automation and contribute to higher flexibility. In this type of deployment, multiple NFV vendor products are managed and orchestrated on a single commodity NFVI by a single NFV MANO entity. This architecture is the telecom equivalent to cloud computing.
}

\textbf{Our contribution:} In this paper, we present a Proof-of-Concept (PoC) for creating an overlay network that ensures the confidentiality of data passing between VNFs in a 5G network slice with the integration of Open Source MANO (OSM), which is one of the dominant service orchestrators, and the efficient VPN technology - WireGuard.  %WireGuard shows superior performance in terms of latency and throughput compared to existing VPN technologies. That is the reason for using it in our VPN as a Service (VPNaaS) proof of concept. OSM is used to manage and orchestrate components and their interconnections, which constitutes a more extensive network infrastructure. 
We present the first OSM-WireGuard framework and its lifecycle. The northbound interface of OSM, combined with juju proxy charms, allows us to manage the WireGuard infrastructure with our API easily. The performance results show that the proposed framework is a promising solution for providing traffic isolation in slices with strict latency and throughput requirements. This is the first application and implementation of WireGuard in a 5G network setting where OSM facilitates the configuration and key distribution; thus, it opens new potentials for both technologies.

\section{Background}
\subsection{Technical background} \label{backT}
\textbf{OpenStack}~\footnote{\url{https://www.openstack.org/}} is a cloud operating system providing Infrastructure-as-a-Service (IaaS) orchestration, and fault and service management to operators. In our framework, OpenStack acts as a VIM hosting the base images, referred to as Virtual Deployment Units (VDUs), on which the VNFs are instantiated and the virtual network infrastructure connecting the VNFs. OSM handles the rest of the orchestration, configuration, and management of VNFs by interacting with OpenStack to set up the network services.

\textbf{Open Source MANO (OSM)}~\footnote{\url{https://osm.etsi.org}} is a tool developed by ETSI for the management and orchestration of VNFs.
OSM makes use of Virtualized Network Function Descriptors (VNFDs), which adhere to the unified VNF catalogue, to describe all VNFs and network services in a standard way. VNFs are packed into VNF packages, which include the VNFD, configuration scripts, and other artifacts. Operators can then combine these packages to describe their network service, in a Network Service Descriptor (NSDs), and then again combine NSDs to provide a network slice instance.

In the OSM community~\cite{osmjune2019}, the process of producing a VNF package, making the package functional by satisfying its lifecycle stages, creating network service, and consequently, a network slice is known as the \textit{VNF on-boarding process}. This process consists of three stages, which are known as Day-0, Day-1, and Day-2. They all revolve around the full lifecycle of a single VNF package. 
\begin{itemize}
    \item \textit{Day-0}: Determines all necessary elements of a VNF package for its successful instantiation, and sets up its management, so it is possible to configure the VNF later. The main requirements in this stage are: describing different involved components of the VNF (the essential VDUs for hosting VNFs), designating NFVI requirements, indicating topology and management mechanism for the VNF, stipulating the certain Linux images and cloud-init files, and identifying the instantiation parameters.
    \item \textit{Day-1}: The VNF is instantiated and configured to enable the continuous delivery of its service. It grants the capability to instantiate and initialize the necessary parameters to configure the VNF in order to provide the expected network service. The essential steps in this stage are: classifying dependencies between the involved VNF components and determining the fundamental configuration for service initialization.
    \item \textit{Day-2}: Provides all necessary elements for the VNF package to be fully operational. It gives the possibility to re-configure VNFs' operations and modify them during the runtime process. Day-2 operations include reconfiguration, monitoring of Key Performance Indicators (KPIs), and automatic scaling based on KPIs' status.
\end{itemize}
We use OSM to deploy VNFs, network services, and network slices, and to manage and orchestrate WireGuard. 

\textbf{Juju}~\footnote{\url{https://jaas.ai}} is one of the main VNF Managers (VNFMs) for OSM. %It is running on the OSM and it maintains SSH connection to its VM. 
It works by interacting with charms that act as a structured VNFM for VNFs on the NFVI. A charm holds all the necessary hooks to manage the lifecycle of a VNF, such as deployment, configuration, and exposing services to external processes. 

\textbf{WireGuard} is a new protocol for secure network tunneling proposed in 2017 \cite{wireguard:donenfeld}. It aims to replace IPsec and TLS-based solutions for most use cases. WireGuard demonstrates that it is possible to implement a secure network tunnel, using state of the art cryptography in less than 4000 lines of code. Reported results show that WireGuard outshines both IPsec and OpenVPN in terms of throughput and ping time. 

WireGuard’s approach to key distribution is agnostic and relies on out-of-band mechanisms. 
It operates solely on layer 3 and works by associating public keys with IP addresses, and it identifies peers strictly by their public key. We explain it with a simple example, taken from \cite{wireguard:donenfeld}.

\begin{figure}[!htb]
  \centering
  \includegraphics[width=0.9\linewidth]{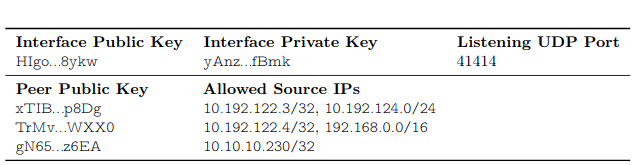}
  \caption{The cryptographic key routing table of a host 1a \cite{wireguard-wp}.}\label{fig:wireguard-conf-1a}
\end{figure}

When the sender transmits a packet out the \texttt{wg0} interface, it consults the cryptographic key routing table, shown in Figure \ref{fig:wireguard-conf-1a}, to determine which public key to use for encryption. For instance, a packet destined to \texttt{10.192.122.4} would be encrypted with the \texttt{TrMv...WXX0} public key. This also works for receiving packets; if the interface \texttt{wg0} receives a packet, after decryption and authentication, it will only accept the packet if the IP address resolves in the table to the public key used in the secure session for decrypting it. This process dramatically simplifies firewall rules from an administrative perspective, because any packet arriving on the WireGuard interface will have a reliable authentic source IP address.
%\vspace{-4mm}

\begin{figure}[!htb]
  \centering
   \subfigure[]
  {\includegraphics[width=0.9\linewidth]{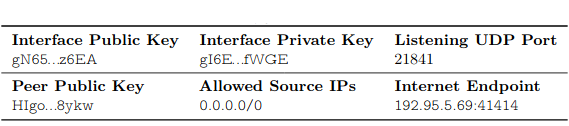}\label{first}}
   \hfill
 \subfigure[]
  {\includegraphics[width=0.9\linewidth]{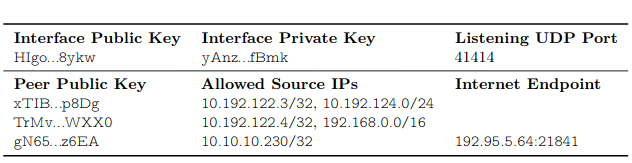}\label{second}}
  \caption{The cryptographic key routing table for a mobile scenario \cite{wireguard-wp}.}\label{fig:wireguard-conf-1b}
\end{figure}

The following example applies to implement a secure network tunnel in a mobile network environment where mobility is a priority. Given the two cryptographic key routing tables shown in Figure \ref{fig:wireguard-conf-1b}, where the peer identified by the public key \texttt{gN65...z6EA} in Figure \ref{first} has configured the cryptographic key routing table to contain an endpoint for the peer \texttt{HIgo...8ykw}, and changed the IP address and port of the WireGuard interface to \texttt{192.95.5.64:21841}. The host \texttt{gN65...z6EA} sends an encrypted packet to \texttt{HIgo...8ykw} at \texttt{192.95.5.69:41414}. When \texttt{HIgo...8ykw} receives the packet, it learns that the endpoint for sending a reply has changed. Thus, it updates the table in Figure \ref{second} and ensures that the reply reaches the new endpoint.
%\vspace{-4mm}

\subsection{Related Work}

In \cite{7846998}, a threat analysis is presented, in the context of NFV based network services and a conceptual design framework for NFV based security management and service orchestration, according to the ETSI-NFV MANO framework. Authors argued that the tenants' lack of visibility and control over the network resources might lead to security threats and vulnerabilities. The tenant only has access to the prepared VNF and not to the image before build time; therefore, it is hard for the tenant to know how data is accessed. Thus, the NFV ecosystem relies on VNF providers that follow best practices and allow as much transparency as possible into their VNF - to avoid security incidents. They concluded that even though there are several ETSI-NFV MANO implementations, few of them focus on NFV security management and orchestration. Their proposed conceptual framework, secMANO, expands the ETSI NFV MANO framework by including security orchestration and policy components. secMANO provides security functions as a service, such as access control to NFV resources. Recently, much has happened in the ETSI-NFV MANO ecosystem. OSM introduced Role-Based Access Control (RBAC), which restricts access to OSM API calls and ensures that only authorized tenants can interact with the NFV resources, which motivates our choice for an orchestrator and the relevance of our proposed framework.

Reference \cite{GUNLEIFSEN201977} identifies the need for per-flow encryption in between chained NFVs and, in this way, to isolate the end-user traffic between VNFs. It presents the Software Defined Security Associations (SD-SA) component as an alternative to the Internet Security Key Exchange protocol (IPsec-IKE). IPsec-IKE is not applicable inside a VNF due to the lack of bidirectional data plane communication channels between chained VNFs. SD-SA automates the dynamic VPNs establishment in a NFV domain by introducing the VNF manager and SDN controller, where the last is omitted from the implementation. In our work, we use the latest technologies for encryption and orchestration. 
%A new architecture of a key exchange mechanism in distributed NFV environments.

The use of WireGuard as a VPN for future industrial systems has been investigated in \cite{8758010}. %The authors compared various VPN technologies, including WireGuard. 
The reported results confirm that WireGuard outperforms IPsec in terms of throughput and latency on different hardware platforms.
Reference \cite{PLAGA2019596} discusses the use of WireGuard as an open-source solution for ensuring secure communications on the transport layer in terms of confidentiality, integrity, and authenticity for future decentralized industrial IoT infrastructures. The automatic establishment of secure authenticated connections, harder DoS attacks, less resource consuming cryptographic primitives, better throughput performance, simple setup, and code of fewer than 4000 lines that also allows formal verification make WireGuard an appealing solution for embedded IoT appliances. %These works highlight WireGuard's relevance in a 5G environment where the low latency impact and high throughput make it a strong contender for future 5G networks. 
These are the key objectives of WireGuard that motivated us to develop a PoC of using WireGuard as a VPN solution in 5G. Private 5G networks tailored for a specific enterprise will certainly foster Industry 4.0 \cite{8377166}. 

%The survey of security in 5G \cite{8712553} identifies six security domains: network access security, network domain security, user domains, application domain, service based architecture, visibility and configurability of security.

%Another direction of security in 5G is optimizing the placement of network functions while taking into consideration the security deployment constraints. The authors of \cite{7592416} focus on the efficient placement of virtual security function when deployed using SFCs, to make sure that they are used most effectively and least expensively. They first transform the high level security needs by the tenants into a set of security patterns, then introduce the security deployment constraints and take them into account together with the deployment costs for security implementation.

A group of NFV vendors showcased how to built a virtualized multi-vendor LTE Packet Core on top of OpenStack and commercial-off-the-shelf servers orchestrated by OSM \cite{whitestack:sc:1}. They realized a site-to-site VPN.
The difference is that we aim to provide a point-to-point solution for VPN. Since VNFs are instantiated in multiple domains of the network architecture, a tenant might want the confidentiality of the traffic between the VNFs secured in a true end-to-end fashion. 

\section{OSM-WireGuard Framework}
\begin{figure}[h]
  \centering
  \includegraphics[width=0.95\linewidth]{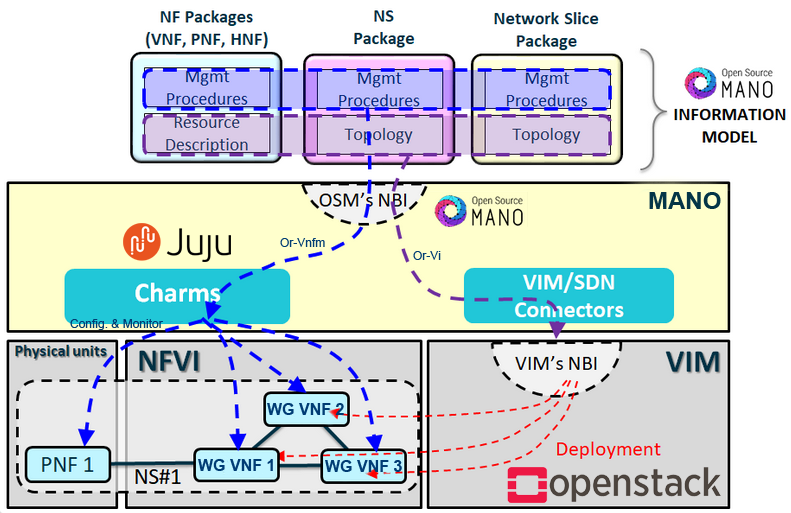}
  \caption{Our proposed OSM-WireGuard framework, where the NBI of OSM makes use of the extensive information model to minimize integration efforts of network operators. Proxy charms are used to configure VNFs deployed on the NFVI \cite{OSM:SF:WP}.}\label{fig:osm-nbi}
\end{figure}

Here we present the main contribution of this paper: the OSM-WireGuard framework where OSM manages and orchestrates services in a network that provides a VPN tunnel between endpoints in a virtualized 5G environment. OSM works by utilizing its extensive information model aligned with ETSI-NFV, which makes it possible to model and automate the lifecycle of VNFs, network services, and network slices. OSM has several ways to support the configuration of WireGuard on each VNF. In the proposed OSM-WireGuard framework, illustrated in Figure \ref{fig:osm-nbi}, the OSM Northbound Interface (OSM-NBI), in combination with Juju proxy charms, manages the WireGuard-enabled VNFs running on NFVI through OSM actions. The use of proxy charms with the OSM-NBI allows script configurations on the VNFs.
The charms make it easy to expose functions to the administrator to define tasks such as key generation, adding and deleting peers, starting and stopping the WireGuard service, and gathering performance metrics. In our case, the administrator uses the OSM-NBI to configure peering relationships between gateways, which allows the RBAC feature to control configuration access of the gateways. The installed VIM is Microstack, a minimal version of OpenStack, with only the core components installed. To run OSM and the proxy charms, we use Microk8s. Note that Microk8s performance satisfies our needs for this first attempt towards integrating OSM and WireGuard, since the launched VDUs in this implementation do not demand very high computational tasks.
The technical specifications of the testbed are listed in Table \ref{tbl:test-2-req}. The commands and files used during the installation and configuration of the testbed can be found in \cite{simen2020}.

\begin{table}[!b]
\centering
\caption{\label{tbl:test-2-req} Hardware and OS used for the Microstack.}
\begin{tabular}[b]{| c | c |}
\hline
Host OS & Ubuntu 18.04 Bionic \\ \hline
Memory & 32 GB  \\ \hline
CPU & Intel i7-4790 \\ \hline
NIC & NetXtreme BCM5722 \\ \hline
\end{tabular} 
\end{table}

\iffalse
\begin{figure}
  \centering
  \includegraphics[width=0.9\linewidth]{test-2-stack.png}
  \caption{The software stack used during the development of the VPN-as-a-Service PoC.}\label{fig:test-2-stack}
\end{figure}
\fi
\subsection{OSM-WireGuard lifecycle}

\begin{figure}
  \centering
  \includegraphics[width=0.85\linewidth]{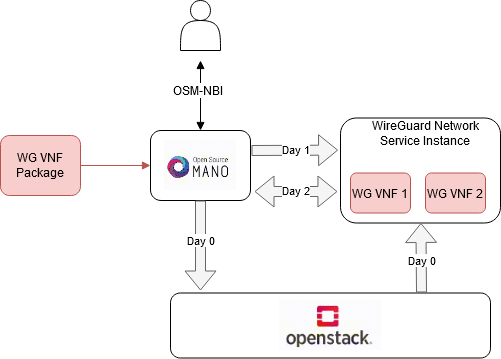}
  \caption{Lifecycle of the WireGuard VPN-as-a-Service solution with OSM.}
  \label{fig:lifecycle}
\end{figure}

Day-1 and Day-2 of the three days lifecycle impact directly the configuration of WireGuard in the network service. Adding VNF configuration with OSM requires defining our configuration primitives in the VNFD before we specify them in detail in our proxy charm. Adding primitives to the VNFD simplifies the process and oversight of the generated API to verify and manage it, as it provides us with a detailed list that can be read as documentation of the configuration primitives and their instantiation parameters. OSM's NBI allows the administrator to configure these primitives when creating the network service through additional YAML files, which automate the process.
A simplified version of the lifecycle is given in Figure \ref{fig:lifecycle}, which portrays the interaction between components in the framework for each day of the lifecycle.
 
\subsection*{Day-0}
The administrator must first retrieve the VNF and network service packages, make sure the hardware requirements are met, upload the packages to the OSM platform through the NBI, and confirm that a compatible Ubuntu image exists on the connected VIMs. The administrator then utilizes the OSM-NBI to launch an instance of the network service and make the VNFs manageable from OSM itself.

\subsection*{Day-1}
During this process, as illustrated with the Day-1 arrow in Figure \ref{fig:lifecycle}, OSM deploys Juju charms to instantiate the network service, according to the parameters given at instantiation time. The process is also depicted in Figure \ref{fig:osm-nbi}. The extra configurations included with \textit{--config} are used to specify instantiation parameters, which make it simpler to effectively launch the service and customize the network service's attributes according to the network service provider's needs.

\subsection*{Day-2}
When the network service is instantiated, and all WireGuard-enabled VNFs are running on their specific VDUs, the network service is ready to be operated at runtime. By this point, the administrator can use the OSM-NBI, and the actions \textit{add-peer} and \textit{del-peer} to manage peering relationships between new WireGuard-enabled VNFs while running Day-2 operations. During this time, the OSM KPI monitoring framework can also be used to monitor the performance of the WireGuard gateways, and if there is a need to enhance the performance, OSM can scale the gateways accordingly.

To further connect the WireGuard-enabled network service with other network services, forming a network slice, the administrator can use the Connection Points and Virtual Link Descriptors, which are part of the Network Slice Template. Through the Connection Points, modularity is achieved, and the WireGuard-enabled network service can easily be integrated with the existing infrastructure to provide confidentiality where needed. 
\shorten{
\textbf{First cycle:} We investigated how Wireguard could be used to create a VPN tunnel between two Virtual Machines and how the WireGuard gateways would be launched and handled by a management node. Thus, we had to create a simple network of VMs, as seen in Figure \ref{fig:test-1-topo-with-tunnel}, where the two VMs WG-West and WG-East act as WireGuard gateways and the management VM has WireGuard installed in it and is connected to both of them. The management node is responsible for generating the configuration files and keys. To distribute the keys and configuration files to each WG gateway, we used SSH, as it would be the same technique as OSM’s proxy charms use to interact with the VNF in the network it is managing and orchestrating. 

\begin{figure}
  \centering
  \includegraphics[width=0.9\linewidth]{test-1-topo-with-tunnel.png}
  \caption{Network topology used during the first design cycle, shows two WireGuard gateways connecting subnet Data-East and Data-West.}
  \label{fig:test-1-topo-with-tunnel}
\end{figure}
}

\subsection{Building a network service}

\begin{figure}
  \centering
  \includegraphics[width=0.85\linewidth]{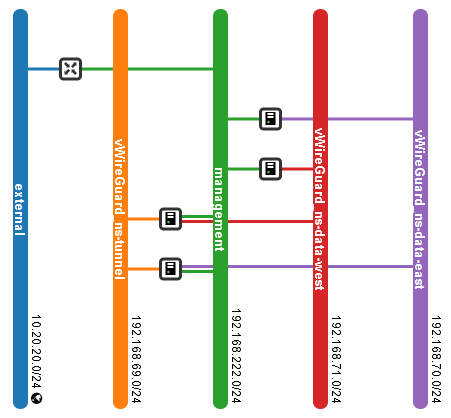}
  \caption{OpenStack's network topology shows the two WireGuard gateways connected to the tunnel network and the test VMs. }\label{fig:test-2-opnstack-topology}
\end{figure}

Here we list the steps of building the VNF and network service packages. 
\begin{enumerate}
    \item Use a minimal Ubuntu Server 18.04 image to create a baseline for the VDU and install WireGuard through cloud-init during the instantiation of the VNF.
    \item Create the VNF package, configure the YAML files, and set path variables in order to create the Juju charm.
    \item Define the list of actions used by the Juju charm framework.
    \item Build the charm, export it to the VNF package, and verify that the VNFD is valid. 
    \item Create the NS package, which includes the required VNFs.
    \item Update the YAML file with the NSD and validate the NSD. 
    \item Compress the packages and upload them to OSM. 
    \item When OSM creates an instance of the network service, the Resource Orchestrator (RO) of OSM instructs the connected VIM to spin up the described NFVI. The outcome of the process is the network topology displayed in Figure \ref{fig:test-2-opnstack-topology}. 
    \item Connect the two WireGuard gateways by adding the peer's public key to the configuration file for the wg0 interface, along with the peer's IP on the VPN, the subnet we want to connect to, and finally, the public endpoint connected top the tunnel-network created by the VIM.
    \item Instantiate the network service. 
\end{enumerate} 

Since network slices are nothing but combined network services in an explicit order, we can see that there are only a few additional steps about configuring the YAML files and the Juju charm framework compared to the steps for network slice instantiation in OSM, as described in \cite{9165419}.
The simplicity and efficiency of WireGuard are apparent as the simple command-line interface makes it easy to understand how most of the process of launching a tunnel network with WireGuard is going to proceed. One thing to note is that with WireGuard’s simplicity, it also becomes a necessity to script or use external tools to solve cases with increased complexity. However, the simplicity is an intended feature to avoid cases such as IPsec, which have myriads of configurations the administrator can choose from - and therefore increase the probability of introducing vulnerabilities through misconfigurations.

\subsection{Key distribution}
Since WireGuard relies on an out-of-band process to perform key distribution, we chose to use Juju proxy charms to handle the creation of peering relationships. During Day-1, the public and private keys of each VNF are generated, while the keys are retrieved through SSH, and distributed with \textit{add-peer} during Day-2. 
\shorten{
As Figure \ref{fig:test-1-key-dist-new} illustrates, the generation of keys and configuration files will be handled by juju proxy charms, which is done after the instantiation of the VM, which constitutes the VNF itself. 

\begin{figure}
  \centering
  \includegraphics[width=0.9\linewidth]{test-1-key-dist-new.png}
  \caption{The sequence-diagram shows the ideal process of generating and distributing keys and configuration files.}
  \label{fig:test-1-key-dist-new}
\end{figure}

A caveat regarding the sequence diagram in Figure \ref{fig:test-1-key-dist-new} is how public keys are distributed throughout the network. There has to be a mechanism on each VM, which pushes the respective public keys to the management node in an asynchronous fashion. Authentication and trust should also be handled at the management level, before distributing keys further onto the network. We make use of OSM-NBI to configure VMs in a VNF. %, it will be the task of the network administrator to distribute the keys. This will lead to a more static approach, which adheres more to the original process of how Wireguard was meant to be used. However, it will be unfeasible to assume that this manual process could be followed for more than a handful of nodes.
}

\subsection{Summary}
\begin{figure}
  \centering
  \includegraphics[width=0.95 \linewidth]{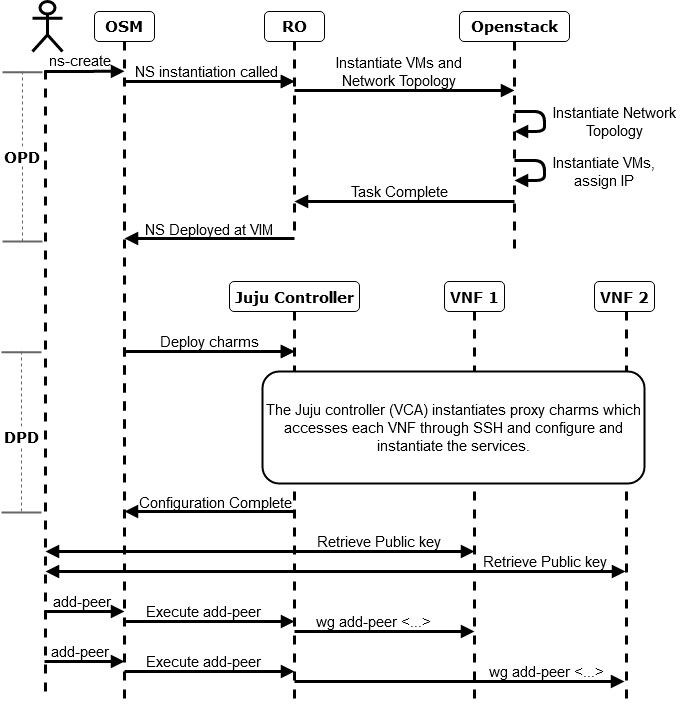}
  \caption{The process of instantiating the WireGuard network service through the use of the OSM-NBI.}\label{fig:initiation-process-design-cycle-2}
\end{figure}

The process of instantiating the WireGuard network service is shown in Figure \ref{fig:initiation-process-design-cycle-2}. The figure also includes some of the key performance indicators (KPIs), On-boarding Process Delay (OPD), and Deployment Process Delay (DPD), which we present in the next Section. %, and it includes the deployment of the underlying infrastructure done by the VIM. 
The process begins when the operator uses the OSM-NBI to call \textit{ns-create}. OSM then logs the instantiation parameters in its database, and the internal RO unit utilizes the OpenStack API to deploy the necessary NFVI. When the deployment of the NFVI is complete, the RO updates the status on OSM's shared message bus, which initiates the configuration of the VNFs. The Juju Controller, which acts as the VNF configuration and abstraction unit, now executes the primitives from the initial-config-primitive section to instantiate the network service. After observing that the VNFs are configured, the operator extracts the public key from each VNF and uses it while calling the \textit{ns-action} \textit{add-peer}, which initiates the process of adding a peer to the VNF's WireGuard configuration. At this point, the network service instance is operational and ready to serve connected network services involved in the network slice. The commands and files for the installation and configuration of the framework can be found in \cite{simen2020}.

\section{Performance Analysis}
One of the goals of 5G-PPP is reducing the average service creation time cycle from 90 hours to 90 minutes~\cite{5gppp_sep_2019}. 
We use and measure the KPIs defined by Yilma et al. in \cite{yilma2019challenges}. On-boarding Process Delay (OPD) is the time it takes to boot-up a virtualized network function image. Deployment Process Delay (DPD) is the time-delay introduced by deploying and instantiating the VNF on the VM to produce an operational network service. 
A summary of the measured OSM-WireGuard framework's KPIs can be found in Table \ref{tbl:nv-mano-kpi}. The service's OPD is 159 seconds, while the DPD ends at 107 seconds. This means that the total time from the point we initiate our VNF image until the instance is ready to provide the required service is 266 seconds. During the on-boarding process, the installation of WireGuard takes 102 seconds, making it the largest contribution to the OPD. The DPD includes the action \textit{add-peer} but excludes the time it takes for the admin to perform ssh connection onto the gateways and extract the public keys - which means that the configuration due to the initial-config-primitives takes 47 seconds.

\begin{table} 
\caption{\label{tbl:nv-mano-kpi} Service creation time.}
\centering
\begin{tabular}{ p{5cm}p{1cm}  }
 \multicolumn{1}{p{5cm}}{NVF-MANO KPI}&\multicolumn{1}{p{1cm}}{Time} \\
 \hline
 On-boarding Process Delay &159 s\\
 Deployment Process Delay &107 s\\
 \hline
\end{tabular}
\end{table}

\begin{table} 
\vspace{-2mm}
\caption{\label{tbl:osm-action-time} Network service action primitives.}
\centering
\begin{tabular}{ p{5cm} p{1cm}  }
 \multicolumn{1}{p{5cm}}{Action}&\multicolumn{1}{p{1cm}}{Time} \\
 \hline
 add-peer & 60 s\\
 del-peer & 51 s\\
 \hline
\end{tabular}
\end{table}

We also compare the throughput and latency of WireGuard and OpenVPN with default settings to determine the benefits of using WireGuard instead of OpenVPN. 
To test the throughput of our network service, we use iperf between the gateway nodes and average the results over 30 minutes. The tests are first performed while WireGuard is active on the two gateways. Then, we stop the WireGuard service, install an OpenVPN server on one gateway, and use the other as a client before running iperf between the nodes again. The tests show that with our configurations, WireGuard has a 5.3 times higher throughput than OpenVPN. 

\begin{figure} 
  \centering
  \frame{\includegraphics[width=0.95 \linewidth]{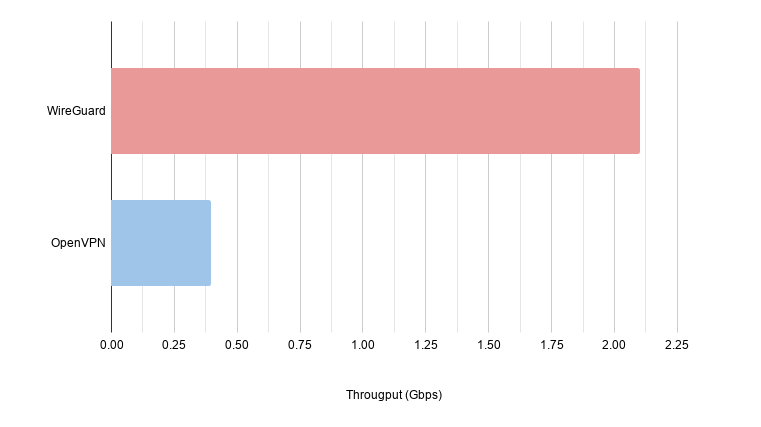}}
  \caption{Throughput of WireGuard and OpenVPN between the gateway nodes averaged over 30 minutes with iperf.}\label{fig:wireguard-vs-openVPN-throughput}
\end{figure}

\begin{figure} 
  \centering
  \frame{\includegraphics[width=0.95 \linewidth]{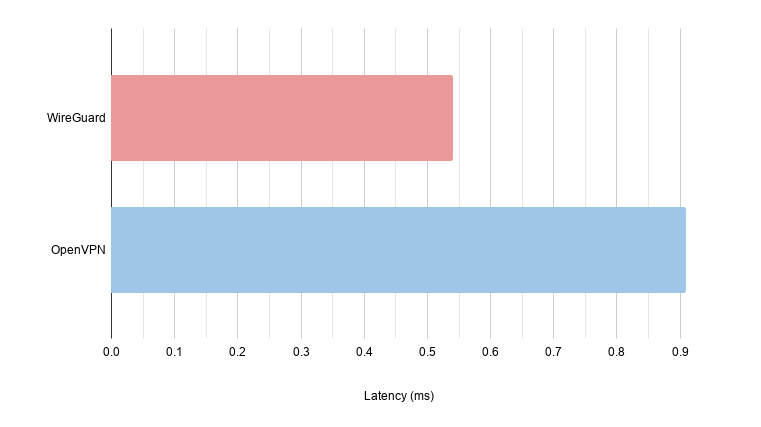}}
  \caption{Average latency, after 1000 ICMP requests between a gateway and a peering subnet's interface on the gateway.}\label{fig:wireguard-vs-openVPN-latency}
\end{figure}

Measuring latency is done by using the ping tool to send 1000 ICMP Requests, and record the average latency between the gateway and a subnetwork on the opposite side of the VPN tunnel. As Figure \ref{fig:wireguard-vs-openVPN-latency} illustrates, packets traveling through the WireGuard VPN have an average latency of 41\% lower than those through OpenVPN.

The performance analysis of the operational KPIs makes it clear that OSM combined with WireGuard can be used to provide VPNaaS while still being in line with the requirement of 90 min for the service creation time. 
The total time that OSM uses to instantiate the network service with two VNFs is 266 seconds, and it can be reduced by about 102 seconds if the operator prepares images used in the VM with a pre-installed and up-to-date version of WireGuard before the on-boarding process. By pre-configuring the images, it is also possible to eliminate actions performed by the Juju charms such as enabling forwarding. Alternatively, the cloud-init API can be used to create the public and private keys on each VNF as well, which would be faster than the Juju charms' additional delay due to their atomic operations. However, as a design decision, it makes more sense to utilize the Juju charms for the generation of keys, as it allows network services that require regeneration of the keys to move the primitive to the standard config primitives, which makes it simple to use the primitive during Day-2 operations. The throughput and latency results of WireGuard and OpenVPN resemble those reported in \cite{wireguard-wp}. 

\section{Conclusions}
We presented the first PoC of providing VPN-as-a-Service in 5G with the use of the WireGuard network tunneling protocol and OSM for management and orchestration of VNFs. The reported performance results show that WireGuard is a more promising protocol compared to OpenVPN, in particular for network slices with low latency and high throughput requirements. Thus, WireGuard should be considered a prime candidate for traffic isolation by means of a VPN.

As future work, we plan to investigate a scenario where the network service or a VNF protected by the WireGuard network service migrates to a new location. In this case, we should use the OpenStack migration features and WireGuard's ability to handle handover. 

\bibliographystyle{IEEEtran}
\bibliography{refer}

\end{document}